\documentclass[%
 aip,
 amsmath,amssymb,
 reprint,%
]{revtex4-1}

\usepackage{graphicx}% Include figure files
\usepackage{dcolumn}% Align table columns on decimal point
\usepackage{bm}% bold math
\usepackage[utf8]{inputenc}
\usepackage[T1]{fontenc}
\usepackage{mathptmx}
\usepackage{etoolbox}

\makeatletter
\def\@email#1#2{%
 \endgroup
 \patchcmd{\titleblock@produce}
  {\frontmatter@RRAPformat}
  {\frontmatter@RRAPformat{\produce@RRAP{*#1\href{mailto:#2}{#2}}}\frontmatter@RRAPformat}
  {}{}
}%
\makeatother
\begin{document}

\preprint{AIP/123-QED}

\title{Extraction of kinetics from equilibrium distributions of states using the Metropolis Monte Carlo method}
% Force line breaks with \\
\author{Sergei F. Chekmarev}
\affiliation{Institute of Thermophysics, SB RAS, 630090 Novosibirsk, Russia }%
%\author{B. Author}%
\affiliation{Department of Physics, Novosibirsk State University, 630090 Novosibirsk, Russia %\\This line break forced with $$\textbackslash\textbackslash
}%
\email{chekmarev@itp.nsc.ru}

\date{\today}% It is always \today, today,

\begin{abstract}
The Metropolis Monte Carlo (MC) method is used to extract reaction kinetics from a given equilibrium distribution of states of a complex system. The approach is illustrated by the folding/unfolding reaction for two proteins - a model $\beta$-hairpin and $\alpha$-helical protein $\alpha_{3}$D. For $\beta$-hairpin, the free energy surfaces (FESs) and free energy profiles (FEPs) are employed as the equilibrium distributions of states,  playing a role of the potentials of mean force to determine the acceptance probabilities of new states in the MC simulations. Based on the FESs and PESs  for a set of temperatures that were simulated with the molecular dynamics (MD) method, the MC simulations are performed to extract folding/unfolding rates. It has been found that the rate constants and first-passage time (FPT) distributions obtained in the MC simulations change with temperature in good agreement with those from the MD simulations. For $\alpha_{3}$D protein, whose equilibrium folding/unfolding was studied with the single-molecule FRET method (Chung et al., J. Phys. Chem. A, 115, 2011, 3642), the FRET-efficiency histograms at different denaturant concentrations were used as the equilibrium distribution of protein states. It has been found that the rate constants for folding and unfolding obtained in the MC simulations change with denaturant concentration in reasonable agreement with the constants that were extracted from the photon trajectories on the basis of theoretical models. The simulated FPT distributions are single-exponential, which is consistent with the assumption of two-state kinetics that was made in the theoretical models. The promising feature of the present approach is that it is based solely on the equilibrium distributions of states, without introducing any additional parameters to perform simulations, which suggests its applicability to other complex systems.  
\end{abstract}

\maketitle

\section{\label{sec:1}Introduction}

The equilibrium distribution of states and the kinetics of their interconversion are two inherent characteristics of any complex system that are important to know for understanding its behavior. At the same time, the former are generally much easier to determine than the latter, both in experiment and in simulations. One representative example is the protein folding/unfolding reaction. Single-molecule experiments, such as the Förster resonance energy transfer (FRET), atomic-force microscopy (AFM) and optical tweezers (OT) techniques, make it possible to determine the equilibrium distributions of states, either in the form of the FRET-efficiency histograms (FRET  \cite{schuler2002probing,huang2009direct,chung2011extracting,liu2012exploring,schuler2013single,%
eaton2021modern}, folding/unfolding) or free energy profiles (AFM \cite{yu2012energy,batchelor2020protein} and OTT \cite{bustamante2020single}, unfolding). However, in order to extract kinetics from these distributions, theoretical models are needed (FRET \cite{mckinney2006analysis,gopich2009decoding,gopich2010fret, ramanathan2015method,osadko2016amplitude}, and AFM and OTT \cite{hummer2001free,hummer2005free,hyeon2007measuring,dudko2008theory}). A similar difference in the possibility of obtaining the equilibrium distribution of states and determining the kinetics exists for the molecular dynamics (MD) simulations of folding of relatively large proteins (of the order of one hundred of residues and larger). While brute force all-atom simulations, which are able to give a detailed insight into the folding kinetics, are very time-consuming \cite{shaw2010atomic,lindorff-larsen2011how}, several methods exist that allow very efficient calculation of the equilibrium distributions of protein states \cite{camilloni2018advanced}, such as the replica exchange molecular dynamics \cite{sugita1999replica,rhee2003multiplexed,rao2003replica} and the weighted histogram analysis method \cite{ferrenberg1989} in combination with sampling from biased molecular dynamics simulations \cite{shea2001from}. Therefore, it is tempting to find out whether the knowledge of the equilibrium distribution of states of a system allows us to obtain information about its kinetics, and if so, to what extent.

The present paper examines the possibility of extracting kinetics from the equilibrium distributions of protein states with the use of the Metropolis's Monte Carlo (MC) algorithm \cite{metropolis1953equation,frenkel2004introduction}, assuming  that the acceptance probability of a new state can be determined by the equilibrium state distribution known from simulations or experiment. The typical examples of such distributions are free energy surfaces (FESs) and free energy profiles (FEPs), although, in general, any equilibrium distribution of states along one or several collective variables can be used for this purpose. In contrast to the nonequilibrium distributions of protein states that determine the probability fluxes of folding \cite{chekmarev2008hydrodynamic,kalgin2014first}, the equilibrium distributions represent potentials of mean-force (PMF) \cite{roux1995calculation,trzesniak2007comparison,shea2001from}, and thus they can be assumed to govern kinetics. A known shortcoming of the MC methods is that a generated MC trajectory represents a Markovian sequence of states rather than the change of these states in time. Therefore, only the relative values of the reaction rates for different reaction conditions, e.g., for temperature, are available. In order to have absolute value of rates, a fit to the time-scale of the original data is required \cite{huitema1999can,rutkai2010dynamic,sanz2010dynamic,bal2014time}. At the same time, in contrast to other possible approaches which could employ the PMFs to calculate kinetics, e.g., the use of Fokker-Planck or Langevin equations \cite{kikuchi1991metropolis,tiana2007use}, no additional parameters are required to perform the MC simulations for a given equilibrium distribution of states. It is a considerable advantage of the present approach. 

The approach is illustrated for two systems. One is a model $\beta$-hairpin protein. In order to have baseline results, the equilibrium MD simulations were performed. The obtained FESs and FEPs were employed as the input distributions of states for subsequent MC simulations. It has been shown that the rate constants of folding and unfolding calculated from the FESs and FEPs change with temperature in good agreement with those from the MD simulations. The other system is an $\alpha$-helical protein $\alpha_{3}$D, for which folding and unfolding rate constants for different denaturant concentrations were extracted from single-molecule FRET trajectories \cite{chung2011extracting} using theoretical models \cite{gopich2009decoding,gopich2010fret}. In this case, the FRET-efficiency histograms (FEHs) obtained in the experiment were employed as the equilibrium distributions of states for the MC simulations. It has been found that the folding and unfolding rate constants obtained in the MC simulations vary with the denaturant concentration in reasonable agreement with those extracted from the FRET trajectories. For both proteins, the first-passage time (FPT) distributions in the MC simulations have been found to be single-exponential, which agree with those obtained in the MD simulations ($\beta$-hairpin protein) and correspond to the assumption of two-state kinetics employed in the theoretical models to extract rate constants ($\alpha_{3}$D protein). 

The paper is organized as follows. Section \ref{sec:2} describes the MD (Sect. \ref{sec:2.1}) and MC (Sect. \ref{sec:2.2}) simulations for the $\beta$-hairpin protein, Sect. \ref{sec:3} present the results of the MC simulations for the $\alpha_{3}$D protein, and Sect. \ref{sec:4} briefly summarizes the results of the work.

\section{\label{sec:2} $\beta$-hairpin Protein: System and Simulation}%

\subsection{\label{sec:2.1}System and Molecular Dynamics Simulations}%

To perform MD simulations, a coarse-grained (C$_{\alpha}$-bead) protein model and a G\={o}-type interaction potential \cite{go1983theoretical} were used. Briefly, the approach is as follows (for details, see Ref. \onlinecite{chekmarev2013protein}). The C$_{\alpha}$-model was constructed on the basis of the solution NMR data for 12-residue HP7 protein (2evq.pdb) \cite{andersen2006minimization}. Specifically, the first conformation in the NMR ensemble of the protein structures was employed for this purpose, which was then considered as a reference native structure of the protein. The G\={o}-type potential accounted for the rigidity of the backbone and the contributions of native and non-native contacts \cite{hoang2000molecular}. Two C$_{\alpha}$-beads were considered to be in native contact if they were not the nearest neighbors along the protein chain and had the interbead distance not longer than $d_{\mathrm{cut}}=7.5${\AA}. In this case, the number of native contacts is $N_{\mathrm{nat}}=N_{\mathrm{nat}}^{\mathrm{NAT}}=27$. The simulations were performed with a constant-temperature MD based on the coupled set of Langevin equations \cite{biswas1986simulated}. The time-step was $\Delta t=0.0125\tau$, where $\tau$ is the characteristic time. At the length scale $l=7.5\mathrm{\AA}${\AA} and the attractive energy $\epsilon=2.2$ kcal/mol \cite{miyazawa1996residue}, $\tau=(Ml^2/\epsilon)^{1/2} \approx 2.6$ ps, where $M=110$ Da is the average mass of the residue. The friction constant $\gamma$ in the Langevin equations was $\gamma=10M/\tau$. In what follows, the length, in particular, the radius of gyration, is measured in angstroms. Unless otherwise noted, the other quantities (energy, temperature and time) are dimensionless; specifically, the energy and temperature are measured in units $\epsilon$ (in the latter case with the Boltzmann constant set to unity), and the time is measured in units of $\tau$. 

\begin{figure}\centering%
\resizebox{1.0\linewidth}{!}{ \includegraphics*{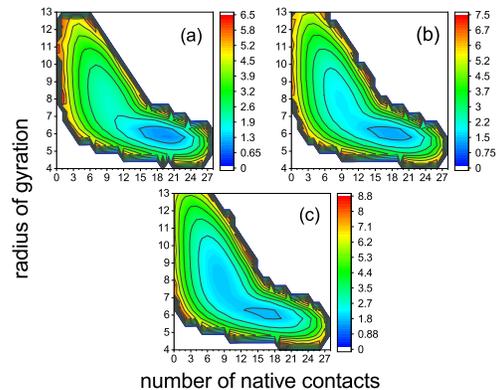}}%
\caption{$\beta$-hairpin, free energy surfaces: ({\bf{a}}) $T=0.35$, ({\bf{b}}) $T=0.4$, and ({\bf{c}}) $T=0.45$. The radius of gyration is given in angstroms.}
\label{beta_fes}
\end{figure}

\begin{figure}\centering%
\resizebox{0.7\linewidth}{!}{ \includegraphics*{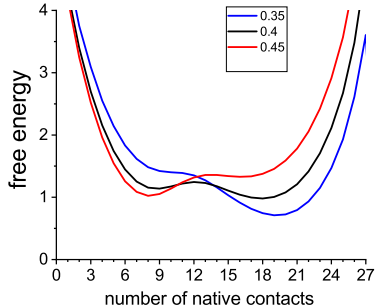}}%
\caption{$\beta$-hairpin, free energy profiles: $T=0.35$ (blue and cyan), $T=0.4$ (black and gray), and $T=0.45$ (red and magenta). The first colors in the brackets are for the MD simulations, and second colors are for the MC simulations based on the FESs of Fig. \ref{beta_fes}. }
\label{beta_fep}
\end{figure}

The equilibrium simulations were carried out for a range of temperatures at which both folding and unfolding events occurred quite frequently, specifically, at $T=$ 0.35, 0,375, 0.4, 0.425, and 0.45. The regime of folding varied from downhill (low temperatures) to uphill (larger temperatures) folding through a marginal two-state folding (intermediate temperatures). For each temperature, the MD trajectory was run until $10^3$ folding/unfolding events took place. The protein was considered to be unfolded if $N_{\mathrm{nat}} \leq 5$, and to be folded if the root-mean-square-deviation (RMSD) from the reference native structure was less than 1{\AA} \cite{chekmarev2021first}. Based on the simulated trajectories, the FESs and FEPs were calculated.  The FES was constructed using the number of native contacts ($N_{\mathrm{nat}}$) and the radius of gyration ($R_{\mathrm{g}}$) as collective variables \cite{best2013native,shea2001from}
\begin{equation}\label{eq1}%
F(N_{\mathrm{nat}}, R_{\mathrm{g}})=-T \ln P(N_{\mathrm{nat}}, R_{\mathrm{g}}) 
\end{equation}
where $P(N_{\mathrm{nat}}, R_{\mathrm{g}})$ is the probability to find the protein in the point $(N_{\mathrm{nat}}, R_{\mathrm{g}})$. The FEP was calculated as 
\begin{equation}\label{eq2}%
F(N_{\mathrm{nat}})=-T \ln P(N_{\mathrm{nat}}) 
\end{equation}
where $P(N_{\mathrm{nat}})$ is the probability that the protein has $N_{\mathrm{nat}}$ native contacts. The resulting FESs and FEPs for the lower, intermediate and upper temperatures are shown in Fig. \ref{beta_fes}{\bf{a}-\bf{c}} and Fig. \ref{beta_fep}, respectively. In structure, they are common for folding of $\beta$-hairpins (experiment \cite{munoz1997folding}, and simulation studies in explicit \cite{zhou2001free,gao2015correct} and implicit \cite{dinner1999understanding,zagrovic2001beta} solvents).   

Both folding and unfolding reactions were considered. For this, the equilibrium MD trajectory was divided into segments of the trajectory corresponding to the transition from the unfolded to the native state (folding) and to the backward transition (unfolding) \cite{chekmarev2021first}. Based on these segments of the trajectory, the MD rate constants of folding and unfolding, as well as the corresponding FPT distributions, were calculated (see below).  

\subsection{\label{sec:2.2} Monte Carlo Simulations}%

Having the FESs and FEPs, the MC simulations were performed. According to the Metropolis algorithm \cite{metropolis1953equation,frenkel2004introduction}, the acceptance probability of the transition from state A to B is determined as
\begin{equation}\label{eq3}%
  W(\mathrm{A} \rightarrow \mathrm{B})= 
    \begin{cases}
      1
    &  {\text{if}} \,\,\,\, P(\mathrm{B}/P(\mathrm{A}) \ge 1
    \\
      P(\mathrm{B})/P(\mathrm{A}) 
      & {\text{if}} \,\,\,\, P(\mathrm{B})/P(\mathrm{A}) < 1
    \end{cases}
\end{equation}
where $P(\mathrm{A})$ and $P(\mathrm{B})$ are the equilibrium probabilities of the old ($\mathrm{A}$) and new ($\mathrm{B}$) states, respectively. A trial move from $\mathrm{A}$ to $\mathrm{B}$ is accepted if $P(\mathrm{B})/P(\mathrm{A}) \ge \alpha$, where $\alpha$ is a random number uniformly distributed between 0 and 1. In application to the free energy landscapes of Figs. 1  and 2, $P(\mathrm{B})/P(\mathrm{A})=\exp\{-[F(\mathrm{B})-F(\mathrm{A})]/T\}$.  For FESs (Fig. 1), $\mathrm{A}$ and $\mathrm{B}$ represent the points in the $(N_{\mathrm{nat}}, R_{\mathrm{g}})$ space, and for FEPs (Fig. 2), the points in the $N_{\mathrm{nat}}$ space. The corresponding free energies $F(\mathrm{A})$ and $F(\mathrm{B})$ are determined by Eqs. (\ref{eq1}) and (\ref{eq2}). 

In the case of FESs, the MC simulation space represented a rectangle with $0 \le N_{\mathrm{nat}} \le 28$ and $0 \le R_{\mathrm{g}} \le 15${\AA}. It was divided into $28 \times 30$ segments ($dN_{\mathrm{nat}}=1$ and $dR_{\mathrm{g}}=0.5$), forming a mesh of $29 \times 31$ discrete states. The trial moves consisted of random shifts of the point along the $N_{\mathrm{nat}}$ and $R_{\mathrm{g}}$ axes by -1 or +1 mesh step so that the 8 nearest points surrounding the current point were sampled with equal probability. To verify that detailed balance is satisfied, a long ("equilibrium") MC trajectory with many folding/unfolding events was run. It was found that the balance was approximately maintained for each point-to-point transition and was exactly fulfilled for the transitions within the simulation cell, i.e. the number of transitions from the central point of the cell to the eight surrounding points was equal to the number of the backward transitions. Accordingly, the equilibrium MC simulations successfully reproduced the original equilibrium distributions on which they were based (one example is shown in Fig. \ref{beta_fep}). To simulate the process of folding, the trajectories were initiated at point (5, 11.8), representing the unfolded state, and terminated at point (27, 5.78), representing the native state, (Fig. \ref{beta_fes}{\bf{a}-\bf{c}}). For the unfolding trajectories, these points served as the terminal and initial states, respectively. 

In the case of FEPs, the MC simulations were conducted in the linear $ N_{\mathrm{nat}}$ space with the boundaries $N_{\mathrm{nat}}=0$ and $N_{\mathrm{nat}}=27$. The trial moves represented random changes of the current value of $N_{\mathrm{nat}}$ by -1 or +1, which were taken with equal probability. In equilibrium,  detailed balance was strictly maintained. The folding trajectories started at $N_{\mathrm{nat}}=5$ and terminated at $N_{\mathrm{nat}}=27$, and the unfolding trajectories used these points as terminal and initial state, respectively. 

\begin{figure}\centering%
\resizebox{0.7\linewidth}{!}{ \includegraphics*{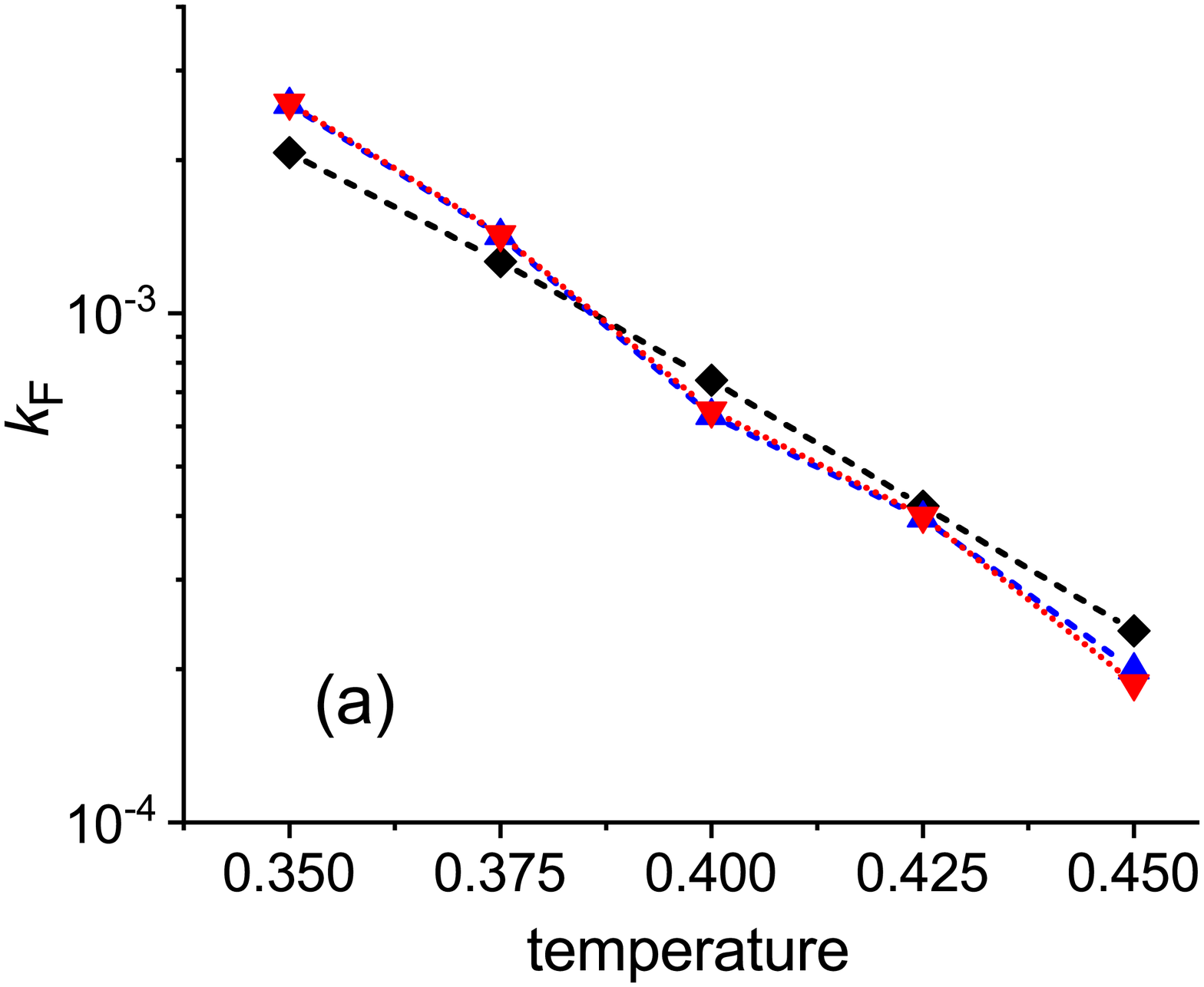}}%
\vfill
\resizebox{0.7\linewidth}{!}{ \includegraphics*{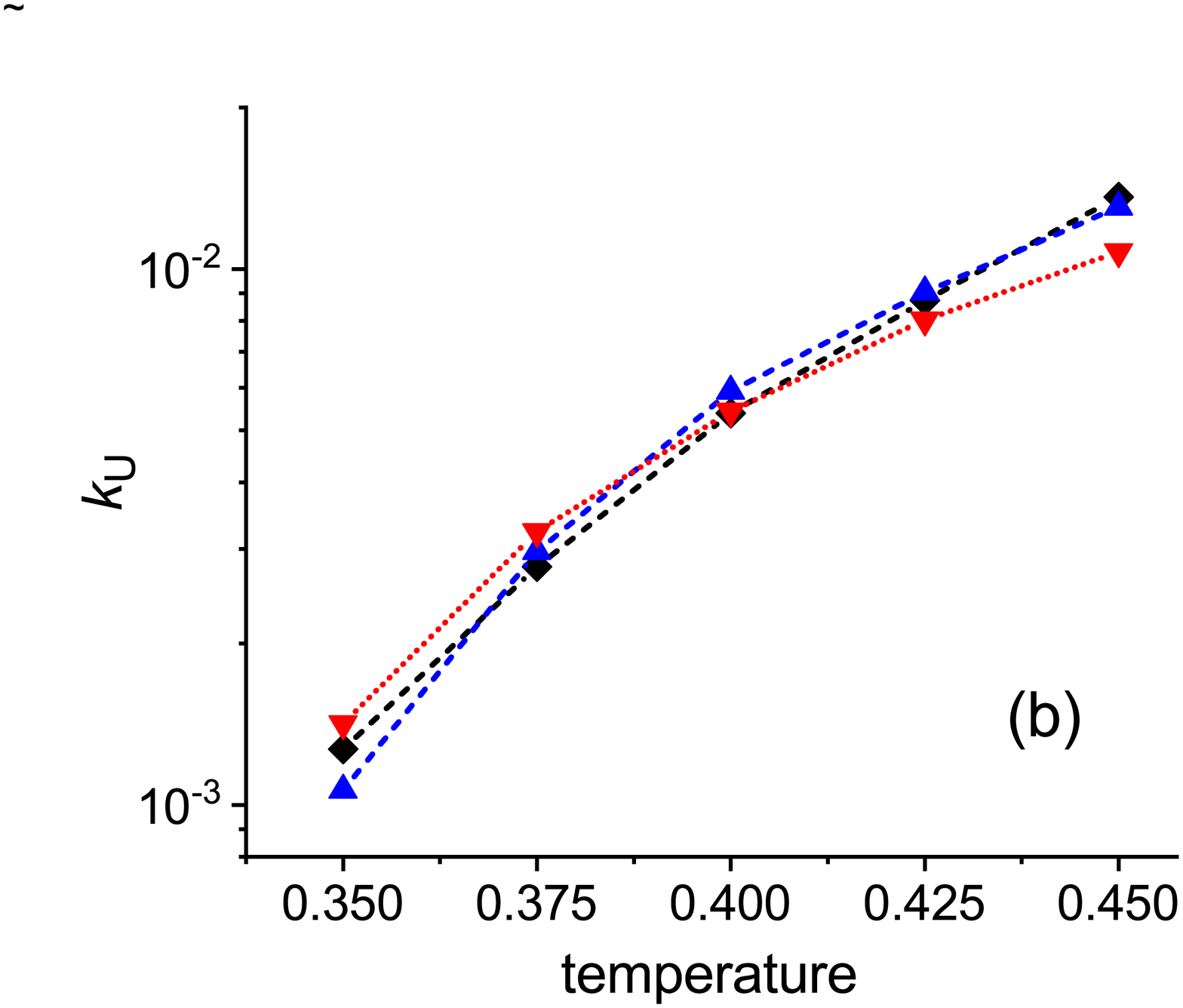}}%
\caption{$\beta$-hairpin, rate constants of ({\bf{a}}) folding and ({\bf{b}}) unfolding: the MD simulations (black diamonds and dashed line), the FES based MC simulations (read triangles and dotted line), and the FEP based MC simulations (blue triangles and dashed line). The lines are to guide the eye. }
\label{beta_k_fold_unfold}
\end{figure}

\begin{figure}\centering%
\resizebox{0.7\linewidth}{!}{ \includegraphics*{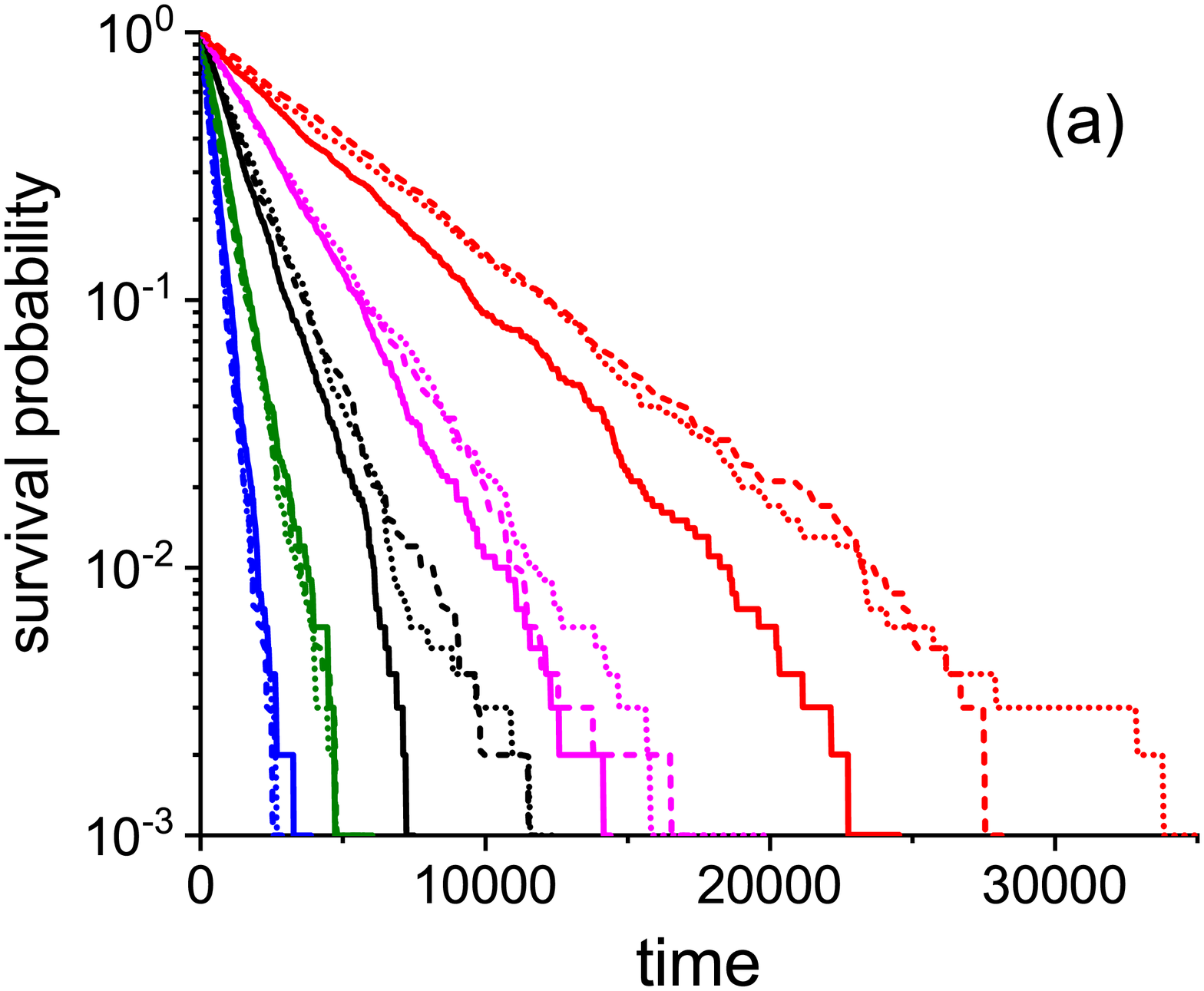}}%
\vfill
\resizebox{0.7\linewidth}{!}{ \includegraphics*{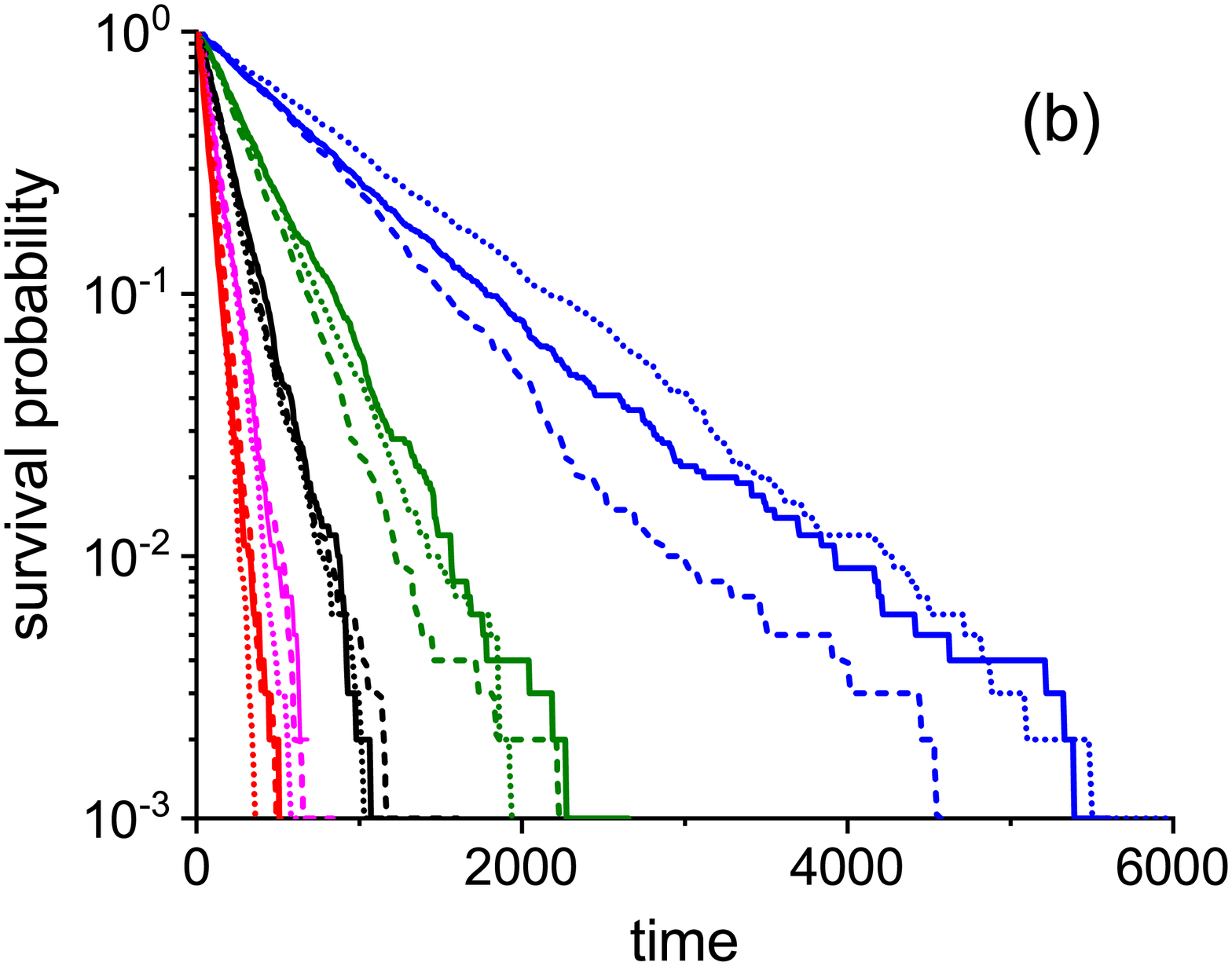}}%
\caption{$\beta$-hairpin, first-passage time distributions: ({\bf{a}}) folding, and ({\bf{b}}) unfolding. Solid lines are for the MD simulations, dashes lines are for the MC simulations based on the free energy surfaces of Fig. \ref{beta_fes}, and dotted lines are for the MC simulations based on the free energy profiles of Fig. \ref{beta_fep}. Temperatures and lines: $T=0.35$ (blue), $T=0.375$ (olive), $T=0.3$ (black), $T=0.425$ (magenta), and $T=0.45$ (red). }
\label{beta_surv}
\end{figure}

In each case (the FES or FEP based simulations, the protein folding or unfolding, and a given temperature), $10^3$ trajectories were run. For the FES based simulations, the acceptance probability increased from $\approx 0.57$ ($T=0.35$) to $ \approx 0.67$ ($T=0.45$) for folding trajectories, and from $\approx 0.57$ ($T=0.35$) to $ \approx 0.62$ ($T=0.45$) for unfolding trajectories, and for the FEP based simulations, it increased from $\approx 0.87$ ($T=0.35$) to $ \approx 0.9$ ($T=0.45$) for both folding and unfolding trajectories.  

In order to fit the MC steps to the MD time steps, the average ratios btween the MC and MD rate constants over the whole temperature range were calculated. In particular, for the FES based folding trajectories, the time step was determined  as $(\Delta t)^{\mathrm{F}}_{\mathrm{FES}} = (1/N)\sum_{i=1}^{i=N}(k_{\mathrm{F}})_{\mathrm{MC},i}/(k_{\mathrm{F}})_{\mathrm{MD},i}$, where $N=5$ is the number of temperatures, and $(k_{\mathrm{F}})_{\mathrm{MC}}$ and $(k_{\mathrm{F}})_{\mathrm{MD}}$ are the rate constants for the MC and MD folding trajectories, respectively. The rate constants were calculated as the inverses of the corresponding mean first-passage times (MFPTs) in units of number of steps (the MC simulations) and $\tau$ (the MD simulations). Similarly, the time steps for the FEP based and unfolding MC simulations were determined. It was found that $(\Delta t)^{\mathrm{F}}_{\mathrm{FES}} \approx 3.4 \times 10^{-3} \tau$, $(\Delta t)^{\mathrm{F}}_{\mathrm{FEP}} \approx 6.3 \times 10^{-3} \tau$, and $(\Delta t)^{\mathrm{U}}_{\mathrm{FES}} \approx (\Delta t)^{\mathrm{U}}_{\mathrm{FEP}} \approx 0.18 \tau$. These figures show that the MC time scales generally depend on both the direction of the process (folding or unfolding) and the dimensionality of the state distributions for the MC simulations (FES or FEP).

Figures \ref{beta_k_fold_unfold} and \ref{beta_surv} compare the results of the MC and MD simulations. Figure \ref{beta_k_fold_unfold} presents the rate constants obtained in the MC simulations in comparison to those in the MD simulations. The MC "time'' scales were adjusted to the MD time scales using the above values of the MC steps in $\tau$ units, e.g., for the MC folding simulations based on the FES, the adjusted rate constant at $T=0.35$ is obtained to be $(k_{\mathrm{F}})_{\mathrm{MC}}= (\tilde{k}_{\mathrm{F}})_{\mathrm{MC}}/ (\Delta t)^{\mathrm{F}}_{\mathrm{FES}} \approx 2.6 \times 10^{-3}$, where $(\tilde{k}_{\mathrm{F}})_{\mathrm{MC}} \approx 8.75 \times 10^{-6}$ is the original rate constant in the inverse number of the MC steps, and $(\Delta t)^{\mathrm{F}}_{\mathrm{FES}} \approx 3.4 \times 10^{-3} \tau$. Figure \ref{beta_k_fold_unfold} shows that the rate constants from the MC simulations are in good agreement with those from the MD simulations through all temperature range. The same result would evidently be obtained if the MC step in $\tau$ units was determined by equating the MC rate constant at one temperature with the corresponding MD rate constant. The above values of the time steps were also used to adjust the time scales for the MC FPT distributions. Figure \ref{beta_surv} presents these distributions in comparison with the distributions obtained in the MD simulations and shows that, similar to the rate constants, they are also in agreement, including a single-exponential decay. Importantly, the good agreement of the MC rate constants with the MD constants through all temperature range evidences that the time step in the MC simulations is not affected by temperature.

\section{\label{sec:3}$\alpha_{3}$D Protein: Experiment and Monte Carlo Simulations}%

\begin{figure}\centering%
\resizebox{0.7\linewidth}{!}{ \includegraphics*{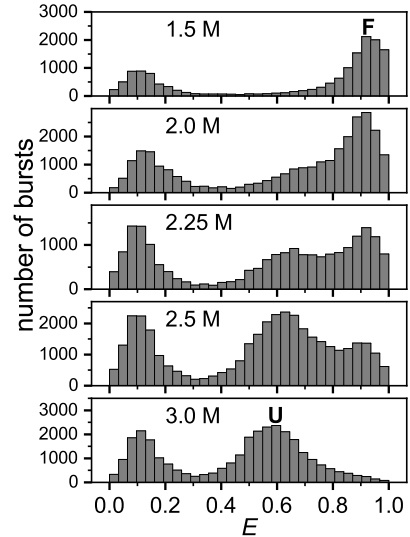}}%
\caption{$\alpha_{3}$D: FRET-efficiency histograms \cite{chung2011extracting}. Figures inside the panels represent the GdmCl concentrations. Labels U and F indicate the unfoded and folded state, respectively.}
\label{a3d_hist}
\end{figure}

The equilibrium folding/unfolding of this 73-residue three-helix bundle protein (2a3d.pdb) at room temperature and different denaturant (GdmCl) concentrations (1.5 M to 3 M) was studied using single-molecule FRET \cite{chung2011extracting}. Two models assuming two-state kinetics were employed to analyze the photon trajectories. One is a Maximum Likelihood Estimation (MLE), in which the photon trajectory were fitted to the sequences of transitions between unfolded and folded states \cite{gopich2009decoding}, and the other is an approximation of the FRET-efficiency histogram (FEH) by a sum of Gaussians whose weights depended on the probabilities of surviving the folded and unfolded states for the bin time (3G) \cite{gopich2010fret}. Using these approaches, the rate constants of folding and unfolding were determined, which were found in good agreement, except for low values of the denaturant concentration, specifically for 1.5 M and, partly, for 2.0 M ({\it cf.} 
the corresponding lines of Table 1 of Ref. \onlinecite{chung2011extracting}).  Figure \ref{a3d_hist} shows the FEHs for freely 
diffusing protein molecules. The time bin to count donor and acceptor photons is 2 ms. Two peaks are significant - at intermediate and higher values of FRET-efficiency ($E$), which represent the unfolded (U) and folded (F) states, as indicated in Fig. \ref{a3d_hist}. The peaks at $E \sim 0$ originate from molecules with inactive acceptors and are not included in consideration. It was found that the FEHs are well approximated by a sum of three Gaussians (3G model). Two Gaussians describe the unfolded and folded states, and the third one accounts for the intermediate values of $E$ that appear due to the transitions between the folded and unfolded states. 
\begin{figure}\centering%

\resizebox{0.7\linewidth}{!}{ \includegraphics*{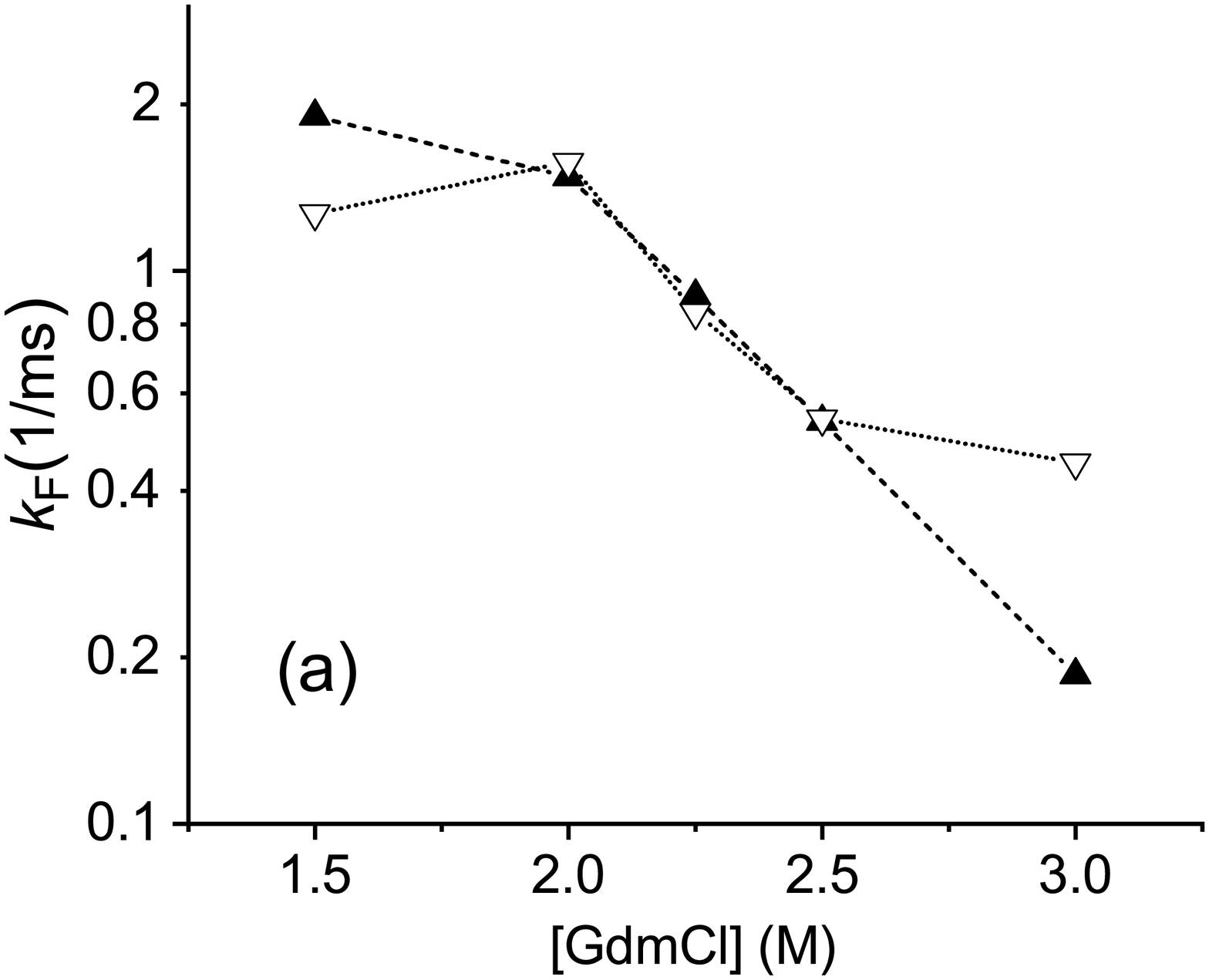}}%
\vfill
\resizebox{0.7\linewidth}{!}{ \includegraphics*{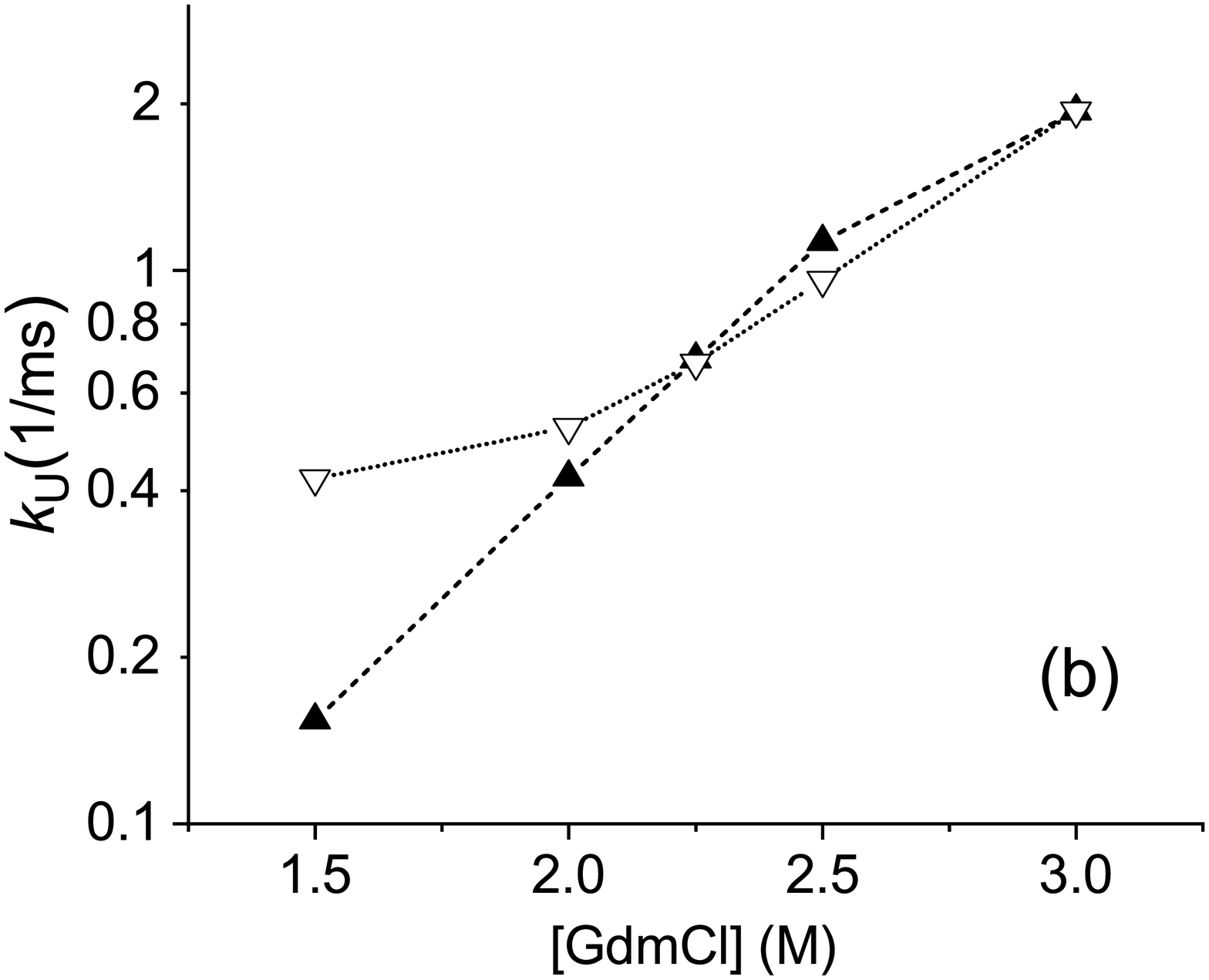}}%
\caption{$\alpha_{3}$D: Comparison of the rate constants for protein ({\bf{a}}) folding and ({\bf{b}}) unfolding. Open triangles show the experimental rate constants obtained with the 3G model \cite{chung2011extracting}, and solid triangles represent simulated rate constants.}
\label{a3d_rates_cmp}
\end{figure}

\begin{figure}\centering%
\resizebox{0.7\linewidth}{!}{ \includegraphics*{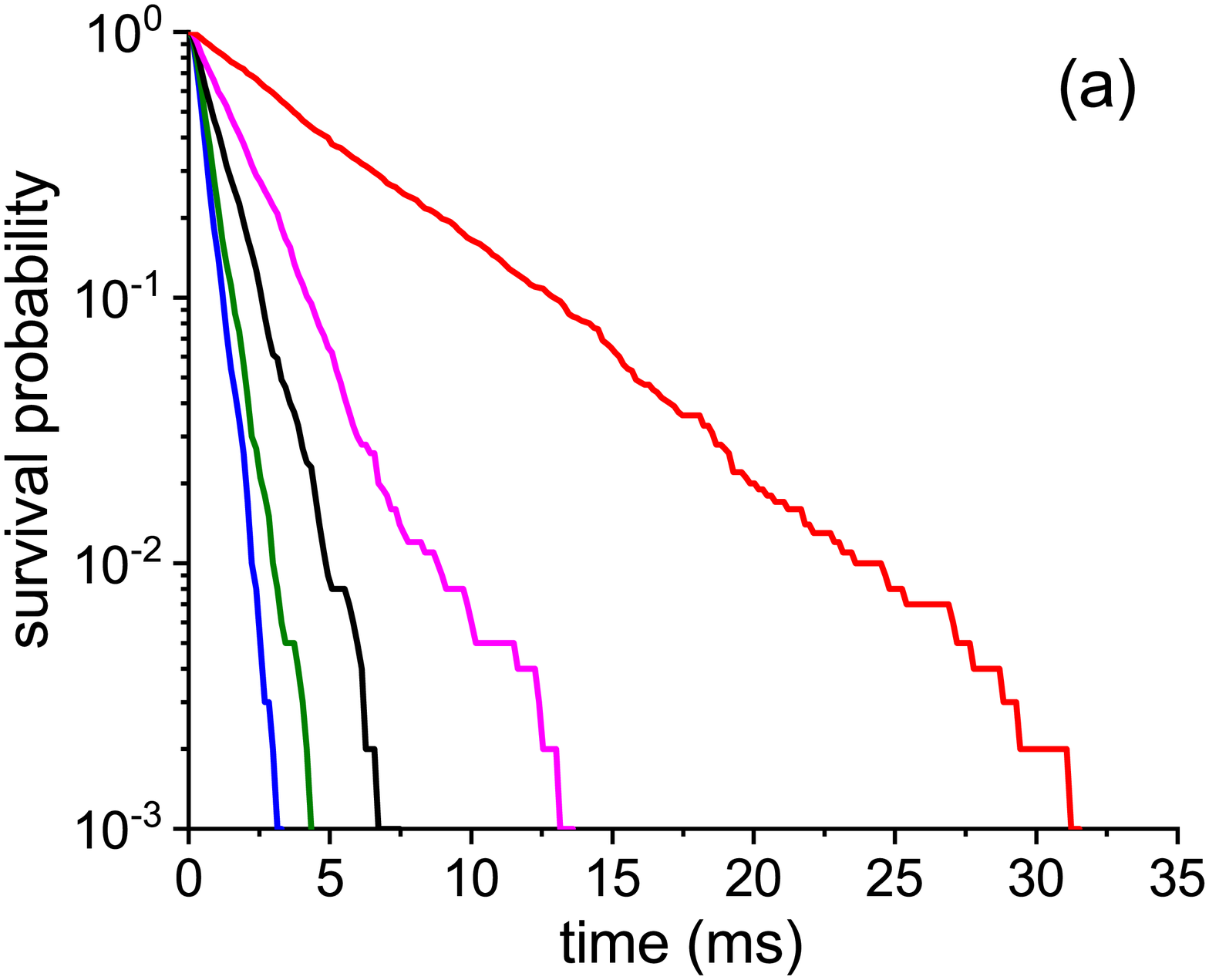}}%
\vfill
\resizebox{0.7\linewidth}{!}{ \includegraphics*{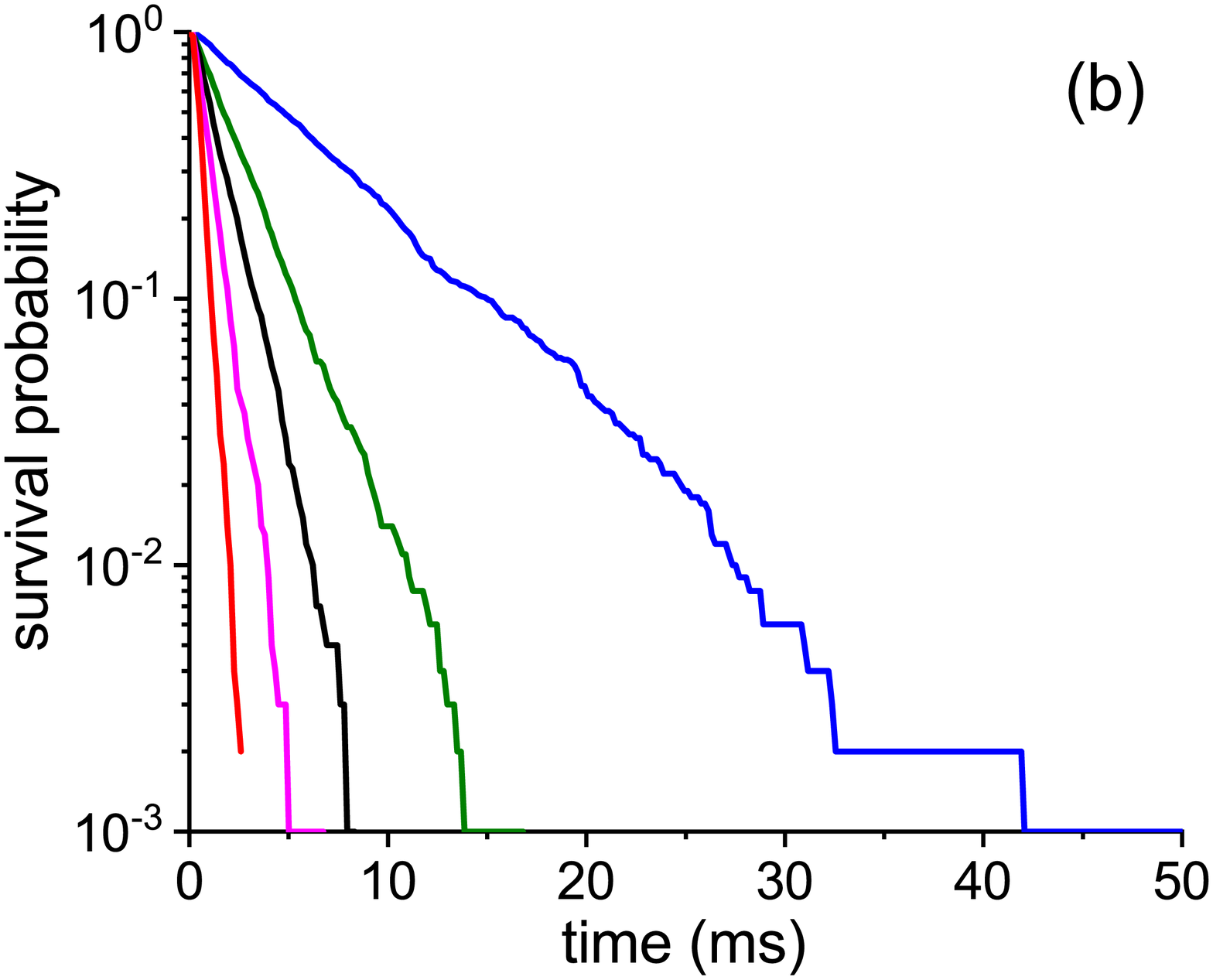}}%
\caption{$\alpha_{3}$D: Simulated first-passage time distributions for protein ({\bf{a}}) folding and ({\bf{b}}) unfolding.}
\label{a3d_surv}
\end{figure}

It should be noted that the present FEHs may not represent the true PMFs, as it would be if they obeyed the F\"{o}rster "distance-efficiency" equation $E=1/[1+(r/R_{0})^6]$ ($r$ is the donor-to-acceptor separation distance, and $R_{0}$ is the F\"{o}rster radius). First, the number of photons in time bin ($\sim 10^2$) is not sufficiently large to avoid the effect of shot noise on the distribution of protein states. In this case, the broadening of the folded and unfolded peaks may mainly be a results of short noise rather than of the presence of folded- and unfolded-like conformations of the protein. Second, the binning time (2 ms) is comparable to the relaxation time ($\sim 1$ms), so that the signal from different protein conformations, which typically interconvert on a much shorter time scale ($\sim 1 \mu$s), may be placed in the same bin. In particular, the intermediate values of $E$ between the unfolded and folded peaks may represent a mixture of photons that come from the unfolded and folded states rather than the transitional conformations. However, if detailed balance is fulfilled, it is more important that the intermediate values of $E$ adequately represent the probabilities of the intermediate protein conformations rather than the conformations themselves.

The MC moves were determined by Eq. (\ref{eq3}), in which the values of FRET-efficiency represented the protein states, and the numbers of bursts in the FEHs served as (non-normalized) probabilities of these states, $P(E)$. According to the FEHs of Fig. \ref{a3d_hist}, the whole range of FRET-efficiencies, from 0 to 1, was divided into 31 segments. To take into account that the peak at $E \sim 0$ corresponds to the inactive acceptor, the populations of states at $0 < E < E_{\mathrm{thr}}$, where $E_{\mathrm{thr}}$ is the value of $E$ at which $P(E)$ has a minimum between the inactive and unfolded peaks, were set to zero, $P(E)=0$. Each MC step was taken to be an equally probable random move to a neighboring segment. Detailed balance was strictly satisfied similar to the MC simulations for $\beta$-hairpin (Sect. \ref{sec:2.2}). The unfolded and folded states were chosen to correspond to the maxima of the unfolded and folded peaks (Fig. \ref{a3d_hist}), with the values of FRET-efficiency $E_{\mathrm{U}}=0.61$ and $E_{\mathrm{F}}=0.93$, respectively. To simulate folding, the MC trajectories were initiated at $E=E_{\mathrm{U}}$ and terminated at $E=E_{\mathrm{F}}$, and to simulate unfolding, the states $E_{\mathrm{F}}$ and $E_{\mathrm{U}}$ were used as the initial and terminal states, respectively. In all cases, i.e., at each denaturant concentration for folding and unfolding, $10^3$ MC trajectories were run. For both folding and unfolding reactions, the acceptance probability varied insignificantly, from $\approx 0.87$ to $\approx 0.92$. Based on the simulated folding and unfolding trajectories, the rate constants for folding ($k_{\mathrm{F}}$) and unfolding ($k_{\mathrm{U}}$) were determined, and the corresponding FPT distributions were constructed. Similar to $\beta$-hairpin, the MC ``time'' scale was adjusted to the experimental time scale by calculating the average ratios between simulated and experimental rate constants. Specifically, the MC time steps for folding and unfolding were determined, respectively, as $(\Delta t)^{\mathrm{F}}=(1/N)\sum_{i=1}^{i=N}(k_{\mathrm{F}})_{\mathrm{MC},i}/(k_{\mathrm{F}})_{\mathrm{exp},i}$ and $(\Delta t)^{\mathrm{U}}=(1/N)\sum_{i=1}^{i=N}(k_{\mathrm{U}})_{\mathrm{MC},i}/(k_{\mathrm{U}})_{\mathrm{exp},i}$, where $N=3$ is the number of denaturant concentrations ranging from 2.0 M to 2.5 M (i.e., 2.0 M, 2.25 M, and 2.5 M), at which the relative populations of both folded and unfolded states were larger than $1 \%$. It was obtained $(\Delta t)^{\mathrm{F}} \approx 7.5 \times 10^{-3}$ms and $(\Delta t)^{\mathrm{U}} \approx 8.7 \times 10^{-3}$ms. Given the MC time steps, the rate constants for folding and unfolding were recalculated as $(k_{\mathrm{F}})_{\mathrm{MC}}= (\tilde{k}_{\mathrm{F}})_{\mathrm{MC}}/ (\Delta t)^{\mathrm{F}}$ and $(k_{\mathrm{U}})_{\mathrm{MC}}= (\tilde{k}_{\mathrm{U}})_{\mathrm{MC}}/ (\Delta t)^{\mathrm{U}}$, where the rate constants denoted by tilde are the original MC constants in the inverse number of the MC steps.

Figure \ref{a3d_rates_cmp}{\bf{a}-\bf{b}} compares the adjusted MC rate constants with those extracted from the FRET-trajectories with the 3G model (Table 1 of Ref. \onlinecite{chung2011extracting}). It shows that the MC simulations based on the FRET-histograms reasonably predict the change of the rate constants with denaturant concentration, for both folding and unfolding reactions. The only exception are the concentrations at which the normalized population of the target state is as small as $\approx 1 \%$, as at 3M GdmCl in the case of folding, where $P(0.93) \approx 1.2 \times 10^{-2}$, and at 1.5M GdmCl in the case of unfolding, where $P(0.61) \approx 9 \times 10^{-3}$. More generally, the MC simulations are expected to give satisfactory results when the peaks for the target states are resolved. Figure \ref{a3d_surv}{\bf{a}-\bf{b}} also presents the simulated FPT distributions for folding and unfolding. The distributions are single-exponential, which is consistent with the assumption of two-state kinetics that was made in the theoretical models for extracting the rate constants from the FRET data  \cite{chung2011extracting}.

\section{\label{sec:4}Conclusions}%

It has been shown that the Metropolis MC simulations based on the equilibrium distributions of states can be successfully used to predict how characteristic reaction times in a complex system change with the determining conditions. As an example, the process of folding and unfolding for two proteins was considered. The first protein, a $\beta$-hairpin protein, whose coarse-grained (C$_\alpha$-bead) model was constructed on the basis of the solution NMR data for 12-residue HP7 protein \cite{andersen2006minimization}, was employed to develop and test the approach. Performing MD simulations, the baseline characteristics of the folding/unfolding process were determined for a set of temperatures - the equilibrium free energy surfaces (FESs) and free energy profiles (FEPs), the rate constants of folding and unfolding, and the first-passage time (FPT) distributions for these reactions. Using the FESs and FEPs as the potentials of mean-force (PMFs), the MC simulations were carried out to obtain relative rate constants of folding and unfolding and the FPT distributions. It has been found that they change with temperature in good agreement with those from the MD simulations and, after fitting the MC steps to the MD time steps, they agree well in absolute value. The other protein is a three-helical $\alpha_{3}$D protein, whose folding/unfolding reaction was studied at different different denaturant concentrations using single-molecule FRET method \cite{chung2011extracting}. In this case, the measured FRET-efficiency histograms (FEHs) were employed as the equilibrium distributions of states. Because of the uncertainty posed by experimental limitations (a relatively small number of photons in time bins, and a long binning time in comparison to the characteristic times of interconversion of protein molecules), it was not quite clear whether these FEHs could be used as the PMFs. However, it has been found that for the denaturant concentrations at which the FEH peak for the target folded (unfolded) state is resolved, the rates constants of folding  (unfolding) change with the denaturant concentration in satisfactory agreement with those extracted from photon trajectories with the use of theoretic models \cite{gopich2009decoding,gopich2010fret}. Also, it has been found the FPT distributions are single-exponential, which is consistent with the assumption of two-state kinetics in the theoretical models  \cite{gopich2009decoding,gopich2010fret} for extracting the rate constants from the FRET data. To obtain absolute values of the rate constants, a time-dependent quantity is required, which, on the one hand, would be known in the experiment (physical or computational) and, on the other, could be obtained in the MC simulations based on the given equilibrium distribution of states, e.g., the self-diffusion coefficient, if this distribution were in physical space \cite{huitema1999can}. The promising feature of the present approach is that it is based solely on the equilibrium distributions of states, without introducing any additional parameters to perform simulations, which suggests its applicability to other complex systems.    

\begin{acknowledgments}
I thank Irina Gopich for valuable discussions of the results of the FRET-experiments for folding of $\alpha_{3}$D protein, and Hoi Sung Chung for sharing the corresponding FRET-efficiency histograms. The work was supported by a contract with the IT SB RAS.\\
\end{acknowledgments}

\section*{Author Declarations}
\noindent
{\bf {Conflict of Interest}}\\
The authors have no conflicts to disclose.

\section*{Data Availability Statement}
\begin{center}
\renewcommand\arraystretch{1.2}
\begin{tabular}{| >{\raggedright\arraybackslash}p{0.3\linewidth} | >{\raggedright\arraybackslash}p{0.65\linewidth} |}
\hline
\textbf{AVAILABILITY OF DATA} & \textbf{STATEMENT OF DATA AVAILABILITY}\\  
\hline
Data available in article or on request from the author
&
The data that support the findings of this study are available within the article or from the author upon reasonable request.
\\\hline
\end{tabular}
\end{center}

\nocite{*}
%\bibliography{aipsamp}% Produces the bibliography via BibTeX.

\end{document}